\documentclass[fontsize=10pt]{IEEEtran}
\IEEEoverridecommandlockouts
\usepackage{graphicx}
\usepackage{epsfig}
\usepackage{color}
\usepackage{subfigure}
\usepackage{tabularx}
\usepackage{amssymb}
\usepackage{latexsym}
\usepackage{amsmath}
\usepackage{color}
\usepackage{relsize}
\usepackage[multiple]{footmisc}
\usepackage{mathtools}
\usepackage{bbold}
\usepackage{soul}
\usepackage{color}
\usepackage{multirow}
\usepackage{xfrac}
\usepackage{xspace}
\usepackage{tikz}
\usepackage{url}

\newcommand*\circled[1]{\tikz[baseline=(char.base)]{
            \node[shape=circle,draw,inner sep=1pt] (char) {#1};}}

\newcommand{\Vijay}[1]{\textcolor{red}{#1}}

\newcommand{\systemName}{\textrm{ZT-RIC}\xspace}
\newcommand{\xAppName}{\textrm{InterClass xApp}\xspace}

\begin{document}
\pagestyle{plain}

\title{\systemName: A Zero Trust RIC Framework for ensuring data Privacy and Confidentiality in Open RAN}

\author{
\IEEEauthorblockN{
Diana Lin$^{*1}$, Samarth Bhargav$^{*2}$, Azuka Chiejina$^{3}$, Mohamed I. Ibrahem$^{4}$, Vijay K. Shah$^{3}$}

\IEEEauthorblockA{$^1$ University of Virginia, VA, USA, $^2$Thomas Jefferson High School for Science and Technology, VA, USA, $^3$ NextG Wireless Lab, North Carolina State University, NC, USA, Augusta University, Augusta, GA, USA}\\

\thanks{$^*$ Equal Contribution} 

$^1$xrc9wg@virginia.edu,
$^2$2025sbhargav@tjhsst.edu,
$^{3}${\{ajchieji, vijay.shah\}@ncsu.edu},
$^4$mibrahem@augusta.edu
}
\maketitle
\begin{abstract}
The advancement of 5G and NextG networks through Open Radio Access Network (O-RAN) architecture marks a transformative shift towards more virtualized, modular, and disaggregated configurations. A critical component within this O-RAN architecture is the RAN Intelligent Controller (RIC), which facilitates the management and control of the RAN through sophisticated machine learning-driven software microservices known as xApps. These xApps rely on accessing a diverse range of sensitive data from RAN and User Equipment (UE), stored in the near Real-Time RIC (Near-RT RIC) database. The inherent nature of this shared, multi-vendor, and open environment significantly raises the risk of unauthorized sensitive RAN/UE data exposure. 
In response to these privacy concerns, this paper proposes a \textit{privacy-preserving zero-trust RIC} (dubbed as, \systemName) framework that preserves RAN/UE data privacy within the RIC platform (i.e., shared RIC database, xApp, and E2 interface). The underlying idea is to employ a computationally efficient cryptographic  technique called \textit{Inner Product Functional Encryption (IPFE)} to encrypt the RAN/UE data at the base station, thus, preventing data leaks over the E2 interface and shared RIC database. 
Furthermore, \systemName customizes the xApp’s inference model by leveraging the inner product operations on encrypted data supported by IPFE to enable xApp to make accurate inferences without data exposure. For evaluation purposes, we leverage a state-of-the-art \xAppName, which utilizes RAN key performance metrics (KPMs) to identify jamming signals within the wireless network. 
Prototyping on an LTE/5G O-RAN testbed demonstrates that \systemName not only ensures data privacy/confidentiality but also guarantees a desired model accuracy, evidenced by a 97.9\% accuracy in detecting jamming signals as well as meeting stringent sub-second timing requirement with a round-trip time (RTT) of 0.527 seconds.

\end{abstract}


\vspace{-0.1in}
\section{Introduction}
\vspace{-0.05in}
The Open Radio Access Network (O-RAN) architecture heralds a transformative shift in cellular communications, featuring an open, programmable, interoperable, and virtualized RAN architecture. This novel architecture supports network flexibility and scalability besides playing a pivotal role in national security by reducing reliance on foreign vendors and driving economic growth through innovation~\cite{whitehouseBrief}. 

At the heart of the O-RAN paradigm is the concept of intelligent and data-driven closed-loop control through the \textit{RAN Intelligent Controller (RIC)} component, specifically the Near-Real-Time RIC (Near-RT RIC) which supports telemetry and closed-loop control across multiple RAN sites from different vendors through the use of third-party software microservices known as xApps. These xApps leverage a diverse array of machine learning (ML) techniques that operate on the stored RAN and User Equipment (UE) Key Performance Metrics (KPMs) data such as Received Signal Strength Indicator (RSSI) and Signal to Interference and Noise Ratio (SINR) in the shared RIC database within the Near-RT RIC to make RAN control decisions such as interference mitigation, scheduling, spectrum sharing and traffic steering as demonstrated in works by authors in \cite{guillem2023SenseORAN, bonati2022intelligent,polese2022colo,chiejina2024system}.


While the O-RAN framework offers flexibility, scalability, and cost-effectiveness for cellular networks, data privacy and confidentiality concerns are raised, particularly regarding the data used by ML-driven xApps for RAN control within the Near-RT RIC. The O-RAN Alliance's Security Working Group (WG11) \cite{O-RAN.WG11.Threat-Model.O-R003-v06.00} has conducted a comprehensive security analysis, identifying various threat models across O-RAN components and interfaces. This analysis underscores specific threats and attack vectors affecting ML-based xApps in the Near-RT RIC. 
Currently, researchers evaluate the vulnerabilities in O-RAN, though concrete solutions remain sparse. A recent study developed a framework to detect protocol attacks and identified vulnerabilities in sensitive RAN and UE data, such as IMEI and IMSI, which can be exploited for attacks like IMSI extraction, posing significant privacy risks \cite{wen20245g}.  Additionally, concerns have been raised about potential ML data poisoning attacks that could manipulate stored KPMs in the RIC database to impair network performance~\cite{chiejina2024system, sapavath2023experimental}. Authors in \cite{jiang2023oztrust} propose a zero-trust security system for the O-RAN environment, featuring an access control module for packet tagging and verification of xApps, alongside a policy management module for control.


Current security approaches, such as Role-Based Access Control (RBAC), traditional encryption, and IPSec enhancements recommended by WG11~\cite{O-RAN.WG11.Threat-Model.O-R003-v06.00} are considered inadequate for ensuring robust data security and implementing the zero-trust paradigm within O-RAN. Issues with these mechanisms include documented vulnerabilities such as credential leaks~\cite{github}, besides the computational challenges posed by traditional encryption methods conflicting with O-RAN's stringent latency requirements~\cite{jiang2023oztrust}. Furthermore, existing methodologies mainly focused on protocol attack~\cite{wen20245g} and ML attacks~\cite{chiejina2024system} and fail to address data privacy leaks/attacks within O-RAN architecture. In conclusion, \textit{there is a pressing need for effective data privacy-preserving solutions for the O-RAN that go beyond access control, policy management, and authentication approach, without compromising the real-time operational demands of O-RAN architecture.}

This paper makes the following key contributions:
\vspace{-0.05in}

\smallskip \noindent $\bullet$ We propose \systemName, a zero-trust RIC framework for ensuring data privacy/confidentiality within O-RAN architecture. \systemName adapts a computationally efficient cryptographic technique, called Inner Product Functional Encryption (IPFE), which encrypts the RAN/UE data before storing it in the RIC database. Next, we \textit{quantize} the ML-based xApp model to support the IPFE procedure such that the model makes inferences without decrypting the data, thus ensuring data privacy and confidentiality, unlike the conventional cryptographic methods that need data decryption.

\vspace{-0.05in}
\smallskip \noindent $\bullet$ To evaluate the \systemName framework, we utilize an example ML-based \xAppName as designed in \cite{chiejina2024system}, which aims to detect the presence of a jammer in a wireless environment using RAN-related key performance metrics. Then, leveraging an over-the-air LTE O-RAN testbed, we show that \systemName framework ensures data privacy while achieving quantized accuracy of up to \textbf{97.9\%} for the \xAppName model. Moreover, for this level of accuracy, the encryption and model evaluation times for the \xAppName model summed up to \textbf{0.474s} and achieved a round trip time (RTT) of \textbf{0.527s}. We observe that the performances of \systemName are at par with that of the baseline O-RAN framework (no data privacy protection) which validates that our proposed \systemName framework addresses privacy issues without negatively impacting network performances and latency requirements. 

The rest of the paper is organized as follows. Section 2 covers O-RAN background, threat model, and design
objectives. Section 3 presents the ZT-RIC framework while Section 4 details the O-RAN testbed and ZT-RIC prototype. Section 5 discusses experimental results, and lastly, Section 6 concludes the paper.

\section{O-RAN Background, Threat Model, and Design Objectives}
\begin{figure}
    \centering  \includegraphics[width=0.8\linewidth]{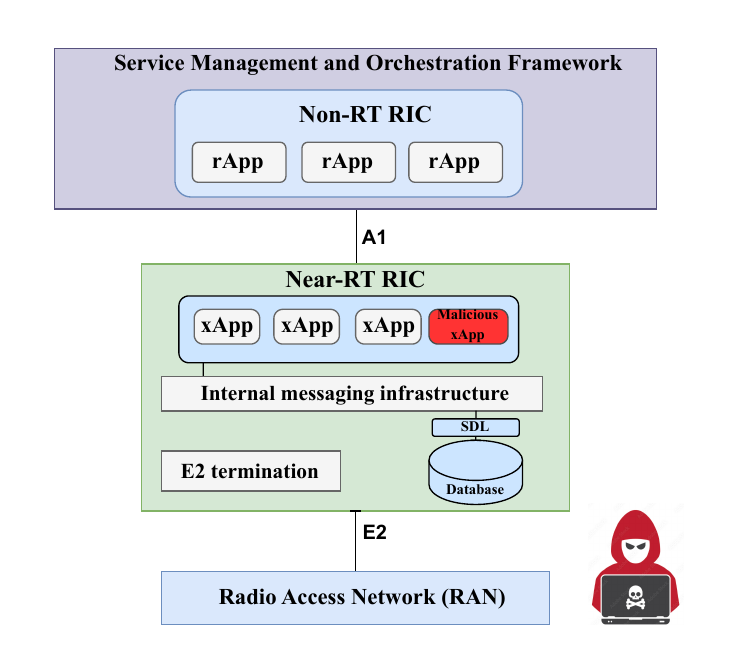}
    \vspace{-0.1in}
    \caption{O-RAN Architecture showing internal and external adversaries. 
    }
    \label{fig:oran}
    \vspace{-0.2in}
\end{figure}

\vspace{-0.05in}
\paragraph{O-RAN Background}
Fig.~\ref{fig:oran} shows a simplified O-RAN architecture highlighting major components and possible internal and external adversaries. For a detailed background of O-RAN architecture, please refer to \cite{abdalla2022toward}. We briefly discuss key O-RAN components critical for our proposed work. 

\smallskip \noindent $\bullet$ \underline{\textit{RAN Intelligent Controller (RIC):}}
The RIC is a pivotal component of the O-RAN architecture, responsible for hosting programmable components that process data and employ ML algorithms to formulate control policies that optimize the RAN. It consists of two main logical RIC controllers, each operating at different time scales. These controllers are:

    \textit{1. Non-real-time (Non-RT) RIC}: The Non-RT RIC resides within the service management and orchestration framework (SMO) and handles control loops at a time granularity of $>1$s. 

     \textit{2. Near-real-time (Near-RT) RIC:} The Near-RT RIC operates control loops between 10ms to 1s. It hosts third-party vendor applications called \textbf{xApps}. These xApps act as \textit{intelligent components} and run ML algorithms that are used to determine control policies for optimizing the RAN through the E2 interface. Other major components of the near-RT RIC include the RIC database/Shared Data Layer (SDL) and internal messaging infrastructure which helps to connect multiple xApps and also ensures message routing. 


\smallskip \noindent $\bullet$ \underline{\textit{RIC Database:}}
The RIC database functions as a centralized, shared repository, playing a crucial role in storing a diverse range of data. Its primary purpose is to store RAN/UE data that can be shared with various coexisting xApps. The RIC database is instrumental in storing metrics related to the communication between RAN nodes and associated UEs. Depending on the chosen split option in the architecture, this data can include I/Q samples/spectrograms or KPMs such as throughput and Signal-to-Interference-plus-Noise Ratio (SINR). The current O-RAN industry-standard adopts a functional split of $7.2$, which enables access to RAN KPMs, and thus, is primarily considered as the RAN/UE data in this paper.

\paragraph{RIC Threat Model against Data Privacy}\label{Threat}
We take into account threats from both internal and external adversaries in the O-RAN system as shown in Fig. \ref{fig:oran}.

 \smallskip \noindent $\bullet$ \underline{\textit{External adversaries:}} The open interface, such as E2 between the RAN and RIC, may be attacked by a malicious intruder, e.g., an eavesdropper, who could intercept the sensitive RAN and user-exchanged data to learn their data to obtain confidential and sensitive information about them.
 
\smallskip \noindent $\bullet$ \underline{\textit{Internal adversaries:}} In addition to providing their functionalities and services honestly, the RIC components such as xApps may attempt to access data or learn sensitive information about users and RAN or even other xApps’ data. As identified in the technical report by O-RAN Alliance WG11, one of the key attack vectors is that the RIC database may leak raw and sensitive RAN and UE-specific data to malicious xApps or third-party software microservices co-hosted by the RIC platform~\cite{O-RAN.WG11.Security-Near-RT-RIC-xApps-TR.0-R003-v03.00}. This will potentially allow an intruder to export sensitive data from the near-RT RIC database, monitor the database for specific information, and take adverse action. In Fig. \ref{fig:oran}, we show a possible internal adversary in the form of a malicious xApp.

\paragraph{Design Objectives}
The aim of this work is to achieve the following functionality and data privacy goals.

 \smallskip \noindent $\bullet$ \textbf{Privacy requirements.} No entity, including RIC xApps or other microservices and O-RAN E2 interface, should be able to access or learn the sensitive RAN/UE data (e.g., KPMs), within the O-RAN architecture. 
    
 \smallskip \noindent $\bullet$ \textbf{Functionality requirements.} In O-RAN architecture, the proposed privacy-preserving framework should allow the benign/trusted xApp to utilize the RAN and UE data (stored in the RIC database) for the intended RAN control decision-making. In addition, it must adhere to the O-RAN's Near-RT RIC timing requirements of \textit{$<1$ second round-trip time} while preserving xApp's model inference accuracy and cellular network performance.

\paragraph{ML-based \xAppName}\label{sec:xAppDetails} To investigate the data privacy issues as well as evaluate the performance of the proposed solution, we utilize a state-of-the-art \xAppName~\cite{chiejina2024system} as an exemplary xApp. While we only test \systemName with \xAppName, the overall functionality and analysis remains the same across any xApp that leverages \systemName. \xAppName aims to detect the presence of a jammer transmitting an interference signal in the operating cellular environment. The xApp utilizes a multi-layer perceptron (MLP) model architecture that consists of five layers including the input and output layers. The input layer has $M$ neurons, and the output layer has $2$ neurons. The hidden layers have $30$, $15$, and $7$ neurons, respectively. These hidden layer structures are determined after tuning the hyperparameters based on training and validation purposes. 

The xApp utilizes four KPMs, namely, bitrate, MCS, BLER, SINR, and BSR, that represent input features to the MLP model. We consider (m) KPMs and collect (t) different time windows of metrics to stack together and form an extended array of input before feeding into the MLP. Feeding multiple time windows into the model leads to better accuracy for jammer detection compared to feeding a single time window to the model \cite{chiejina2024system}. For training, we considered $m = 5$ and $t = 10$, which means the number of input nodes, $M$ is $50$ $(10 \times 5)$ in total resulting in an input shape of $(50,1)$. The model has a total of $2123$ parameters. The hidden layer and the final output respectively have a ReLU and softmax activation function. 
\vspace{-0.12in}
\section{Privacy-preserving \systemName Framework}
\begin{figure}
    \centering  \includegraphics[width=1\linewidth]
    {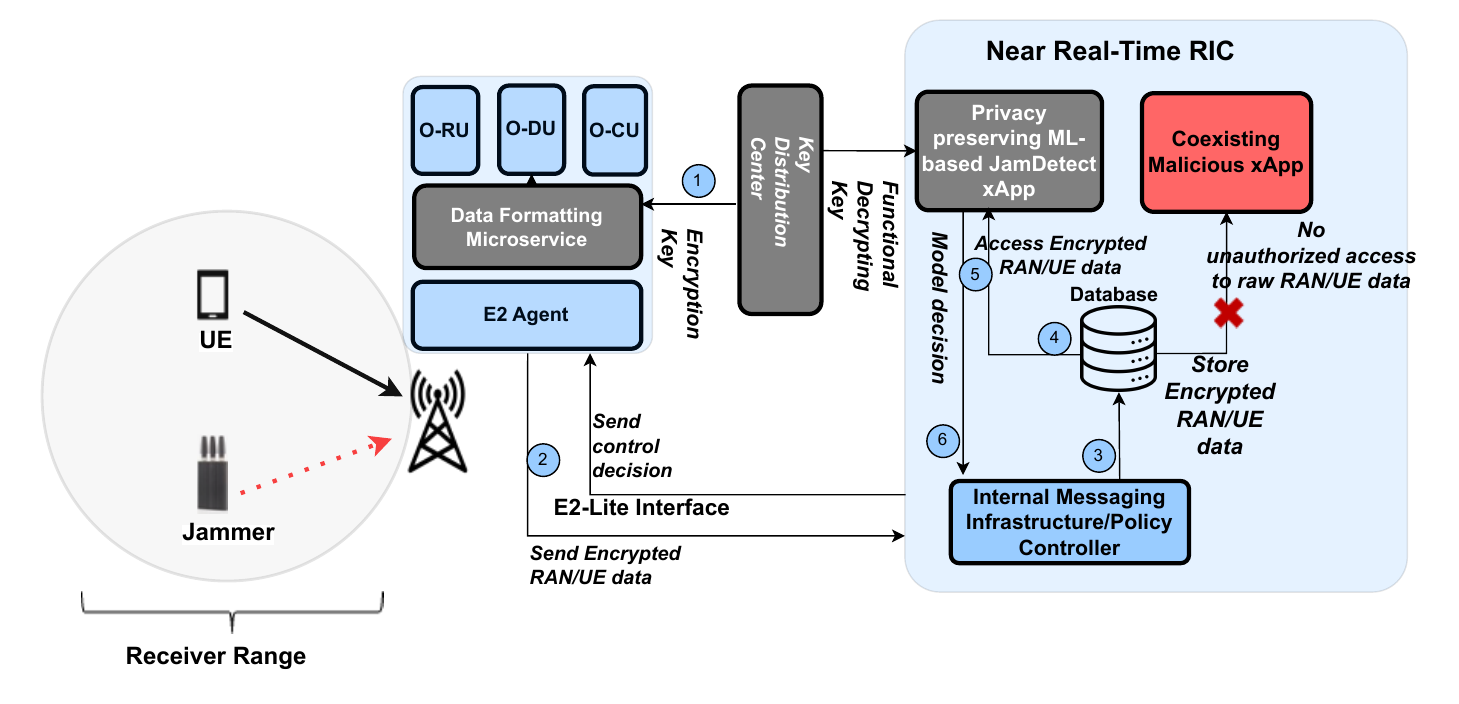}
    \vspace{-0.25in}
    \caption{Overview of \systemName Framework.}
    \label{pporansteps}
    \vspace{-0.2in}
\end{figure}

\vspace{-0.08in}
Fig. \ref{pporansteps} shows an overview of the proposed privacy-preserving \systemName framework. \systemName introduces three new/modified functional blocks, namely, \textit{key distribution center (KDC)}, a \textit{privacy-preserving \xAppName} (as an example xApp) and an \textit{encryption microservice} (Notice grey boxes in Fig. \ref{pporansteps}). 
The step-wise flow of the \systemName framework is outlined below.
\vspace{0.05in}

\noindent \textbf{Step \circled{1}} The KDC is responsible for generating and distributing the \systemName keys, including, the encryption key to the RAN and the functional decryption key to the xApp. The \textit{encryption key} is used to encrypt the RAN/UE data, and the \textit{functional decryption key} is used for executing the ML-based xApp on the encrypted data and getting the classification result only.

On the RAN side, a microservice is introduced which serves the purpose of processing and encrypting the RAN/UE data before they are forwarded to the near-RT RIC. The encryption is done using the keys derived from the KDC. \textit{Note that our proposed framework requires that the data processing followed by encryption of the raw RAN/UE data are implemented at RAN itself, thus, the raw data is never exposed to the near-RT RIC thereby protecting against internal/external adversaries.}

\noindent \textbf{Step \circled{2}} After the E2 connection is established between the near-RT RIC and the RAN, the encrypted data is sent to the near-RT RIC via the E2-Lite interface. 

\noindent \textbf{Step \circled{3}} The IMI forwards the encrypted data to the RIC database for xApps (including, our our example \xAppName as well as potentially, malicious xApp) to query.

\noindent \textbf{Steps \circled{4} and \circled{5}} Our proposed \textit{privacy-preserving} \xAppName (with quantized ML model elaborated in Section III.b.1) queries the database to retrieve the stored encrypted data, and by using the functional decryption key received from the KDC, it can execute the xApp on the encrypted data without being able to learn or decrypt the original data to \textit{ ensure that the privacy of the data is preserved in the entire data pipeline from the RAN to the legitimate xApp}.

\noindent \textbf{Step \circled{6}} The decision of the model is forwarded to the IMI which then forwards the decision to the RAN for control to optimize the network performance in the presence of a jammer.

\paragraph{A primer on IPFE and motivation behind its incorporation within \systemName} \label{MBIPFE}
IPFE is a type of functional encryption (FE) based public-key cryptosystem, where the possession of a functional decryption key enables learning a specific function $f(\mathbf{x})$ of the plaintext data vector $\mathbf{x}$ given its corresponding ciphertext $\mathbf{C(x)}$ without compromising $\mathbf{x}$~\cite{10.1007/978-3-662-46447-2_33}. Specifically, in IPFE, given a vector $\mathbf{y}$, the ciphertext of another vector $\mathbf{x}$, i.e., $\mathbf{C(x)}$, and a functional decryption key $\text{sk}_{\mathbf{y}}$ associated with a vector $\mathbf{y}$, the inner product result between vectors $\mathbf{x}$ and $\mathbf{y}$, i.e., $\left< \mathbf{x}, \mathbf{y} \right>$, can be determined without compromising or learning the individual values of $\mathbf{x}$. This means that IPFE allows an entity to perform the inner product operation on an encrypted data vector. The following three major entities are involved in the IPFE cryptosystem:

 \vspace{-0.05in}
\begin{itemize}
    \item \textbf{\textit{Key Distribution Center (KDC)}}: The KDC distributes the encryption key to encryptor
    and the functional decryption key $\text{sk}_\mathbf{y}$ to the decryptor.
    \item \textbf{\textit{Encryptor}}: Using the encryption key provided by the KDC, the encryptor can compute a ciphertext of the data vector $\mathbf{x}$ to be sent to the decryptor.
    \item \textbf{\textit{Decryptor}}: Using the ciphertext provided by the encryptor along with the $\text{sk}_\mathbf{y}$ provided by the KDC, the decryptor can compute only the inner product result $\left< \mathbf{x}.\mathbf{y}\right>$.
\end{itemize}

The incorporation of IPFE within the \systemName is motivated by the primary operations in the first layer of the ML-based xApp model being the inner product between the input KPM data vector $\mathbf{x}$ and each column vector $\mathbf{w}_i$ in the weight matrix $\mathbf{W}$ of the first layer, where $\mathbf{w}_i$ is the $i^{th}$ column of $\mathbf{W}$ and hence, $\mathbf{W}$ can be represented by $[\mathbf{w_1},\mathbf{w_2},\dots,\mathbf{w_n}]$ given $n$ columns in $\mathbf{W}$ (assuming the first hidden layer comprises $n$ neurons). This operation can be expressed as $\mathbf{Z} = \mathbf{x} \cdot \mathbf{W}$. Therefore, upon receiving the ciphertexts of the data $\mathbf{x}$, the xApp then employs the $\text{sk}_{\mathbf{W}} = [\text{sk}_{\mathbf{w}_1}, \text{sk}_{\mathbf{w}_2}, \dots, \text{sk}_{\mathbf{w}_n}]$ to derive $\mathbf{Z}$ with $n$ elements which are then added to the first layer's bias vector $\mathbf{b}$ to compute the output of the first hidden layer of the ML-based xApp model. This output serves as input for subsequent layers in the model so their operations can be continued until reaching the output of the last layer without having access to the original data.

In essence, the \systemName framework encrypts data, such as KPMs, before being transmitted to the shared RIC database using the encryption key. In achieving this objective, we adapted the IPFE scheme~\cite{10.1007/978-3-662-46447-2_33} to enable the xApp to securely utilize the encrypted data for its execution without accessing or learning sensitive RAN/user data, compromising xApp performance, or violating the latency requirements of the O-RAN system. Since data only needs to be encrypted once for all xApps to make their predictions, the computational overhead required on the RAN and the memory footprint in the RIC database is also reduced from that of traditional encryption systems. 

\paragraph{\systemName Details}
\label{component}
We detail the three key aspects of the \systemName, namely, \systemName Initialization,  RAN/UE data encryption, and Privacy-preserving xApp inference.

\underline{\textit{1. \systemName initialization:}}
The particular IPFE scheme used in the \systemName framework is inspired from \cite{10.1007/978-3-662-46447-2_33}, and is implemented using the CiFER functional encryption library detailed in~\cite{10.1007/978-3-030-29959-0_1}. For the initialization phase, we undergo two phases:

\vspace{-0.05in}
\smallskip \noindent $\bullet$  \textbf{\textit{Key Generation}}: A key requirement for this step is a trusted key authority, i.e., KDC in our \systemName framework. Note that, such an authority can be operated by a national authority such as the National Telecommunications and Information Administration (NTIA) or a network vendor. It is responsible for generating the system parameters. Specifically, the KDC runs the ${Setup}$ function that outputs $(\mathbb{G}, p, g)$, where $\mathbb{G}$ is a group of order $p$ with generator $g$. It uses these system parameters to generate a master public key ($mpk$) and a master secret key ($msk$) as follows. First, vector $\mathbf{s}$ with a length $l$ is generated, where $\mathbf{s} = [\mathbf{s}_1, \dots,\mathbf{s}_l] \leftarrow \mathbb{Z}_p^l$, where $\mathbb{Z}_p$ is finite field of order $p$. Next, the $mpk$ and $msk$ keys can be computed by performing the following computations.
\begin{align}
    \text{mpk} &= (\mathbb{G}, \mathbf{h}_j = g^{\mathbf{s}_j})_{j \in \left[l\right]} 
\end{align}
\vspace{-0.1in}
\begin{align}
    \text{msk} &= \mathbf{s} 
\end{align}

Then, the KDC computes $n$ functional decryption keys corresponding to each column $i$ of $\mathbf{W}$ by executing the key derivation function $\text{KeyDer}(msk, \mathbf{W})$. This function takes the $msk$ and $\mathbf{W}=\{\mathbf{w}_i\}_{\forall{i}\in n}$, where $\mathbf{w}_i = [\mathbf{w}_i[1], \dots, \mathbf{w}_i[l]] \leftarrow \mathbb{Z}_p^l$, and returns the functional decryption key $\text{sk}_\mathbf{W}$ by performing dot product operations between $\mathbf{w}_i$ and $\mathbf{s}$ as follows.
\begin{equation}\label{SK_eq}
{\text{sk}_\mathbf{W}} = {[\text{sk}_{\mathbf{w}_{i}}]}_{\forall{i}\in [n]} = {\left<\mathbf{w}_i.\mathbf{s}\right>}_{\forall{i}\in [n]}    
\end{equation}
\vspace{-0.05in}
\smallskip \noindent $\bullet$ \textbf{\textit{Quantization}}: Recall that IPFE solely operates on integers, not decimal numbers, both the model and input are transformed into an acceptable result with an input of only integers. Furthermore, IPFE decryption time increases exponentially with larger numbers. Due to the strict time restraints imposed by the specifications of the Near-RT RIC, it would be optimal to limit these integers to a small range. To combat these two issues, we leveraged quantization to transform our trained model into a model that computes results with a representation of the model \cite{PyTorch_Quantization} that operates with integers from 0–255. For each weight, bias, and input, we performed the following transformation~\cite{DBLP:journals/corr/abs-2106-08295}:
    \begin{equation}\label{Quantize_eq}
        \text{Output} = [\frac{\text{Input}}{\text{Scale}} + \text{ZeroPoint}],
    \end{equation}    
where $[.]$ denotes the nearest integer operation. The scales and zero points can be different for each weight layer, bias, and input to each layer. Furthermore, the scales and zero points should be calibrated to work well with the typical input data, providing similar outputs for each layer of the model. In the case of the \xAppName, this typical input data would be a sample KPM. We found that the optimal quantization parameters to accurately represent the sample KPM were $\text{Scale} = 0.079$ and $\text{ZeroPoint} = 0$. By observing how the sample KPM was processed through each layer, we also found a proper scale and well-functioning zero point for every node in every layer.
    

Quantization on its own has a weakness in that the range of input numbers must be approximated through a much smaller, discrete range of numbers. When the range is large, the precision of these approximations quickly decreases. To compensate for this, we "fused" our linear and ReLU layers together in the \xAppName model. Recall that ReLU layers return 0 for any negative input, meaning they reduce the range of possible represented numbers to only natural numbers. By combining the linear and ReLU computations, we preemptively remove any negative numbers from the range of quantized values, thereby increasing the precision of the quantized outputs.


After quantization, the neural network is now compatible with the IPFE protocol. Thus, the KDC can pre-compute $\text{sk}_{\mathbf{W}}$ to be sent to the \xAppName.







\underline{\textit{2. RAN/UE data encryption:}} 
The RAN/UE data is encrypted using the $mpk$ before being sent over the E2 interface and then stored in the RIC database by performing the following operations using the $\text{Encrypt}$ function. The $\text{Encrypt}(mpk, \mathbf{x})$ function uses $mpk$ to encrypt the KPM data vector $\mathbf{x}$, where $\mathbf{x} = [\mathbf{x}_1, \dots, \mathbf{x}_l] \in \mathbb{Z}_p^l$, and outputs its corresponding ciphertext $\mathbf{C(x)}$ by selecting a random number $r \leftarrow \mathbb{Z}_p$, and then computing the ciphertext as follows.
\begin{equation}\label{Enc_eq}
 \mathbf{C(\mathbf{x})} = (g^r, (h_j^r \cdot g^{x_j})_{j \in \left[ l \right]}),
\end{equation}
where $\mathbf{h}_j = g^{\mathbf{s}_j}$ and $\mathbf{C(\mathbf{x})}$ contains two components; $\text{c}_0 = g^r$ and $\mathbf{c} = (h_j^r \cdot g^{x_j})_{j \in \left[l\right]} \in \mathbb{Z}_p^l$.

\begin{figure}
    \centering
    \includegraphics[scale=0.35]{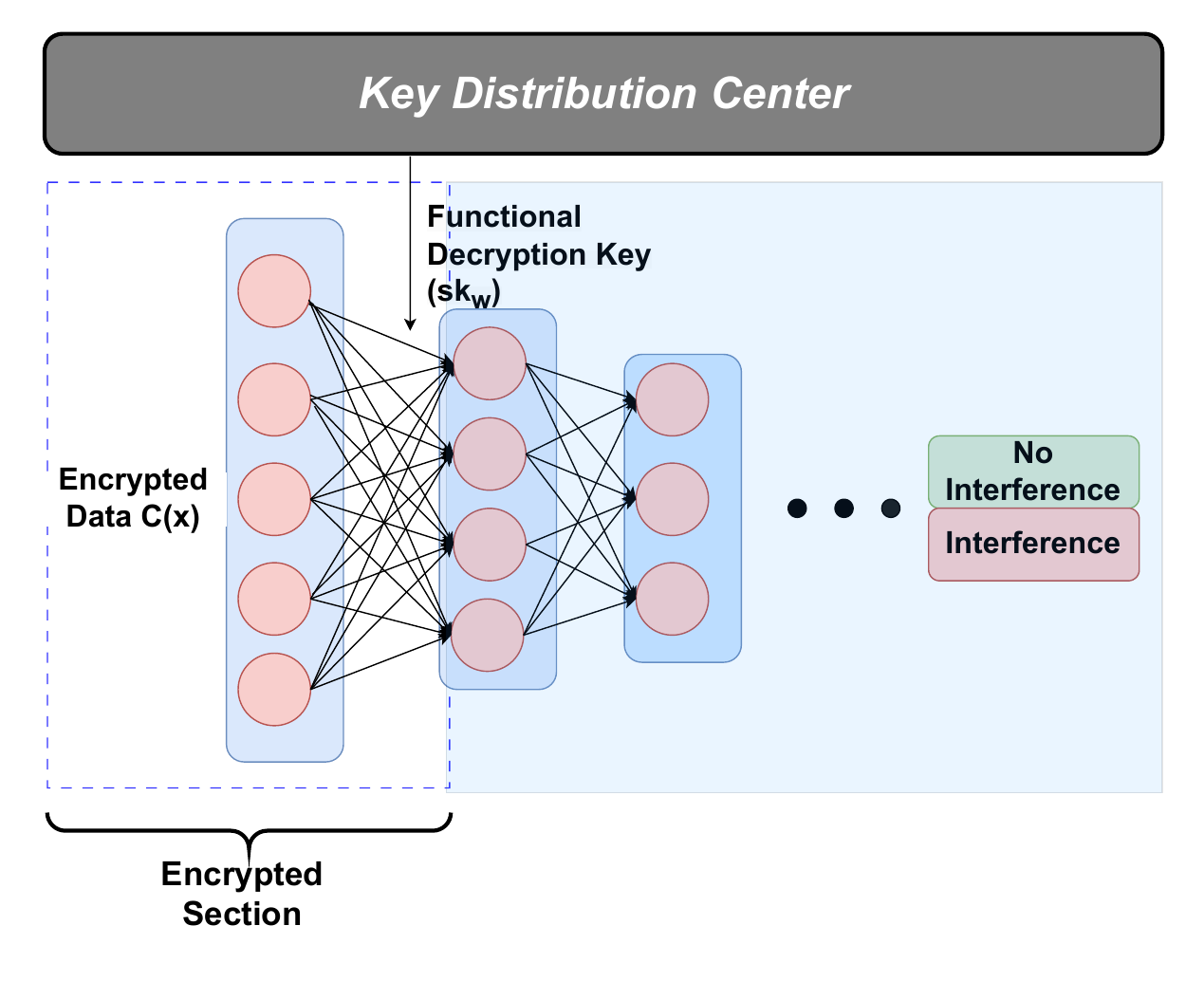}
    \vspace{-0.2in}
    \caption{Illustration of encrypted and non-encrypted parts of the (ML-based) \xAppName within \systemName framework. 
    }
    \label{fig:ML_Model}
    \vspace{-0.25in}
\end{figure}

\underline{\textit{3. Privacy-preserving xApp inference:}}
Lastly, we design a privacy-preserving ML-driven xApp by leveraging the inner product operations of IPFE on encrypted data vectors to enable the xApp to execute the ML-based \xAppName inference model without disclosing or decrypting the data to preserve the RAN/user privacy, as discussed in Section~\ref{MBIPFE}. 

In particular, after receiving the encrypted KPM data $\mathbf{C(\mathbf{x})}$, the \xAppName uses the functional decryption key $\text{sk}_\mathbf{W}$ to make inferences (type of interference in case of considered \xAppName xApp) on the encrypted KPM data. 
The secure evaluation of the \xAppName is done as follows.


\smallskip \noindent $\bullet$ We first run $\text{Decrypt}(\mathbf{C(\mathbf{x})}, \text{sk}_{\mathbf{w}_i})$ for $i \in [n]$, which takes the encrypted input vector $\mathbf{C(\mathbf{x})}= (\text{c}_0, \mathbf{c})$ and the functional decryption key $\text{sk}_{\mathbf{w}_i}$, the xApp can compute the inner product between encrypted input and each column $\mathbf{w}_i$ of the weight matrix $\mathbf{W}$ by performing the following operations. 
\begin{equation}
    \prod_{j \in \left[ l \right]} \mathbf{c}_j^{\mathbf{w}_i[j]} / \text{c}_0^{\text{sk}_{\mathbf{w}_i}} = g^{\left< \mathbf{x}.\mathbf{w}_i\right>}
\end{equation}

Proof of correctness: 
\begin{align*}
\text{Decrypt}&(\mathbf{C(\mathbf{x})}, \text{sk}_{\mathbf{w}_i}) = \frac{\prod_{j \in \left[ l \right]} \mathbf{c}_j^{\mathbf{w}_i[j]}}{\text{c}_0^{\text{sk}_{\mathbf{w}_i}}} \\
&= \frac{\prod_{j \in \left[ l \right]} g^{(s_j r + x_j)\mathbf{w}_i[j]}}{g^{r(\sum_{j \in \left[ l \right]} \mathbf{w}_i[j] s_j)}} \\
&= g^{\sum_{j \in \left[ l \right]} \mathbf{w}_i[j] s_j r + \sum_{j \in \left[ l \right]} \mathbf{w}_i[j] x_j - r(\sum_{j \in \left[ l \right]} \mathbf{w}_i[j] s_j)} \\
&= g^{\sum_{j \in \left[ l \right]} x_j \mathbf{w}_i[j]} = g^{\left< \mathbf{x}.\mathbf{w}_i\right>}
\end{align*}

\smallskip \noindent $\bullet$ These operations are computed for all $\mathbf{W}$'s columns, and hence, the $\text{Decrypt}$ function returns the inner product result between the input KPM data $\mathbf{x}$ and $\mathbf{W}$, i.e., $\left< \mathbf{x}. \mathbf{W}\right>$, by utilizing a discrete logarithm with basis $g$, which can be computed using the Baby-Step Giant-Step algorithm.

\smallskip \noindent $\bullet$ These results, i.e., $\left< \mathbf{x}. \mathbf{W}\right>$, are in clear form and known to the xApp to be added to the bias $\mathbf{b}$ of the first hidden layer to obtain the output of the first hidden layer of the \xAppName ML-based model, which can be represented as follows:
\begin{align*} 
 [(\mathbf{x}\cdot \mathbf{w_1})+\mathbf{b}[1], (\mathbf{x}\cdot \mathbf{w_2})+\mathbf{b}[2], \dots, (\mathbf{x}\cdot \mathbf{w_n})+\mathbf{b}[n]]  
 \end{align*}
Therefore, as shown in Fig.~\ref{fig:ML_Model}, only the operations of the first layer of our model architecture are performed on encrypted data to justify the zero trust paradigm. The result of these inner product operations, i.e., the output of the first hidden layer, is then used as input to the next layer of the model with computations continuing until the output of the last layer is reached, and then getting the classification result. Hence, the \xAppName can evaluate the ML-based inference model without disclosing or decrypting the data to preserve the RAN/user privacy. 
\textit{Note that the number of nodes in the first hidden layer must be smaller than the number of inputs. Otherwise, the xApp may be able to obtain the input private data by constructing and solving a system of linear equations as the number of unknowns is equal to or greater than the number of equations.}

\vspace{-0.15in}
\section{
Standard O-RAN Testbed and \systemName Prototype}
\begin{figure}[!t]
    \centering  \includegraphics[width=1\linewidth, height=3.5cm]{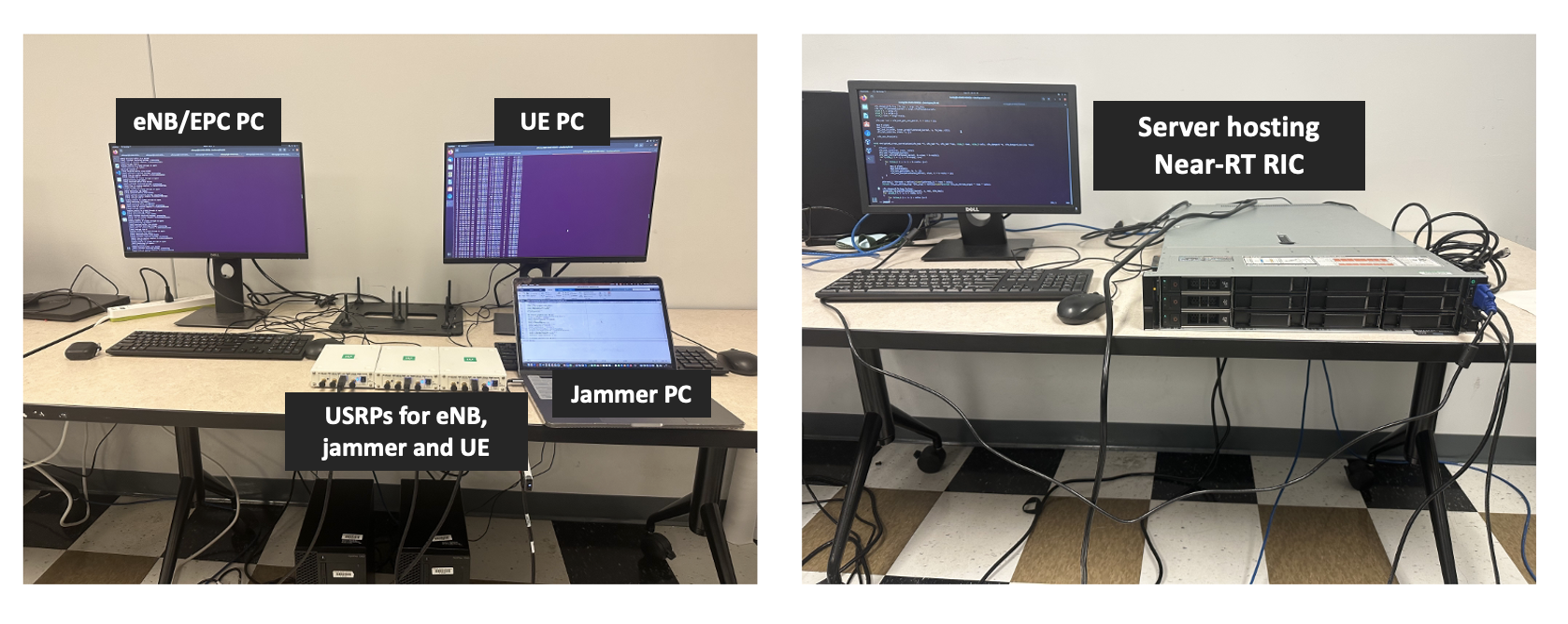}
    \vspace{-0.2in}
    \caption{O-RAN testbed. The left image shows our base station, user equipment, and the jammer USRPs. The right image shows the server hosting the near-RT RIC}
    \label{fig:testbed}
    \vspace{-0.2in}
\end{figure}

\vspace{-0.05in}
\label{testbed}
\paragraph{Standard O-RAN Testbed} 
We leverage an open-source Open AI Cellular (OAIC) platform~\cite{OAIC_website} to build a standard LTE O-RAN testbed. As shown in Fig. \ref{fig:testbed}, the prototyped O-RAN testbed comprises an EPC core, one base station (located on a single compute system), user equipment (UE), a jammer, and a Near-RT RIC similar to the one in \cite{chiejina2024system}.

For the implementation of the RAN/Core and UE, we employed the srsRAN cellular software stack (version 21.10) for creating LTE/5G networks, as noted in \cite{srsRAN}. Each unit operates on Ubuntu release 20.04 OS and is powered by an Intel Core i7-8700 processor, featuring 6 CPU cores, 16GB RAM, 12 threads, and a clock speed of 3.2GHz. The srsRAN stack is known for its adaptability and is ideal for Software-Defined Radio (SDR)-based RANs and UEs. To meet our testbed's specific needs, we modified the srsRAN codebase. These changes include adding a buffer for I/Q sample storage and enhancing RAN control features, like the ability to switch between adaptive or fixed Modulation and Coding schemes (MCS). Both RAN and UE utilize Universal Software Radio Peripheral (USRP) B210 SDRs for managing the radio frequency (RF) front-end tasks, underlining the importance of SDRs in O-RAN's flexible and programmable framework.

The near-RT RIC, responsible for intelligent RAN control, is installed on a rack server that can support multiple RANs. This server is equipped with an AMD EPYC™ 7443P processor, comprising 24 CPU cores, 48 threads, 64GB RAM, and a base clock speed of 2.85GHz. For ease of implementation, we chose an E2-lite interface based on the SCTP protocol, which functions similarly to the E2 interface\footnote{In E2 standard, RAN Functions define specifications and behavior of a service facilitated through E2 interface, and are communicated by the RAN to inform the RIC of its supported capabilities. In the E2-lite interface, no RAN Functions are explicitly communicated by the RAN, which simplifies the connection setup. There is no subscription process and no built-in differentiation of messages between E2-lite nodes.}, and allows exchanging control and report messages between RAN and RIC. 

\paragraph{\systemName Prototype}
We prototype \systemName framework on top of the standard O-RAN testbed. Specifically, we implement the IPFE-based encryption microservice using the state-of-the-art CiFEr C-library~\cite{10.1007/978-3-030-29959-0_1} within the near-RT RIC itself. This is mainly for ease of implementation and should be ideally hosted in the RAN stack (to secure data leakage against external adversaries). Both the encryption microservice and privacy-preserving \xAppName are developed in Python. Thus, in order to facilitate interaction between Python and the CiFEr library (used for IPFE), we utilize the CFFI library~\cite{CFFI_link} to create Python bindings to CiFEr.


\vspace{-0.1in}
\section{Performance Evaluation}
\vspace{-0.08in}
We evaluate the performance of the \systemName framework against the standard (no data privacy protection) O-RAN system in terms of (i) xApp's ML model performance, (ii) network performance, and (iii) timing requirement.

\paragraph{xApp's ML Model Performance:}

We first evaluate the performance of the \textit{quantized ML model} utilized by \systemName's \textit{privacy-preserving} \xAppName  against the baseline MLP model utilized by standard O-RAN system's \xAppName (discussed in Section II.d). Recall that \systemName requires a quantized ML model for IPFE calculation critical for protecting data privacy. The metrics of interest for xApp's ML model performance include \textit{model accuracy} and \textit{false alarms}. 




As shown in Table \ref{tab:KPM Models}, we carry out various experiments with different time windows (default value = $10$) and model parameters (default value = $2123$ parameter) for baseline MLP model (standard O-RAN system) and quantized MLP model (\systemName). Both accuracy and false alarm rate are comparable for both the models, for instance $98\%$ accuracy and $\approx 1.5\%$ false rate for default setting. \textit{This demonstrates that quantizing the MLP model to support the IPFE process (for data privacy) does not negatively impact the xApp's ML model inference accuracy and false alarm rate.}

\begin{table}[t]
    \centering
    \caption{ML model performance for Baseline (Standard O-RAN system) and Quantized ML Model (\systemName)}
   \vspace{-0.15in}
   \scriptsize
    \begin{tabular}{|p{0.3in}|p{0.44in}|p{0.44in}|p{0.44in}|p{0.46in}|p{0.46in}|}
        \hline
         Time Windows & Parameters & (Baseline) Accuracy & (Baseline) False Alarm & (Quantized) Accuracy & (Quantized) False Alarm\\ \hline
         5 & 617 & 97.1\% & 2.9\% & 97.2\% & 2.9\% \\ \hline
         10 & 2123 & 98.0\% & 1.5\% & 98.0\% & 1.4\%  \\ \hline
         20 & 8483 & 97.9\% & 2.0\% & 97.9\% & 2.0\%  \\ \hline
    \end{tabular}
    \label{tab:KPM Models}
    \vspace{-0.2in}
\end{table}
\begin{figure}
    \centering
    \includegraphics[width=0.7\linewidth]{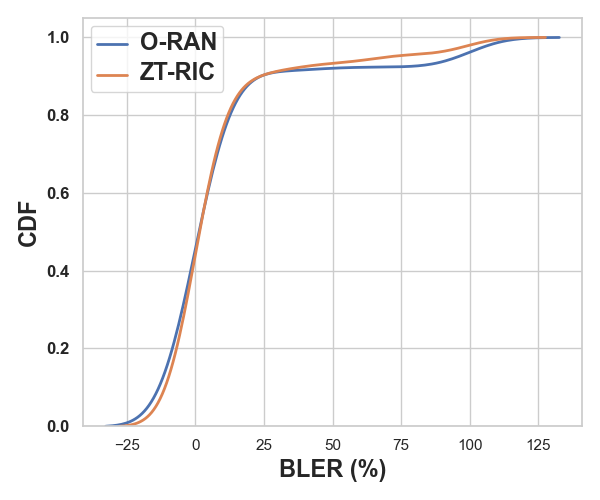}
    \vspace{-0.1in}
    \caption{CDF plot showing BLER performance}
    \label{fig:roc_bler}
    \vspace{-0.15in}
\end{figure}




\paragraph{Network Performance Evaluation}
\label{sec:network}
For network performance evaluation, measured in terms of \textit{Block Error Rate (BLER)}, we use the O-RAN testbed (and \systemName prototype) discussed in Section \ref{testbed}. We initiate an uplink traffic session from the UE to the RAN, lasting $180$ seconds. For the first half of this duration ($90$ seconds), the UE transmits uplink traffic unimpeded by any external disruptions. In the latter half, we introduce over-the-air (OTA) interference from a jammer, exerting a gain of $40$ dB.

Fig. \ref{fig:roc_bler} presents BLER cumulative distribution function (CDF) plot for \systemName against (baseline) O-RAN system. \textit{It can be seen that privacy-preserving \systemName achieves similar network performance as that of baseline O-RAN system.}

\paragraph{\systemName Round Trip Time (RTT)}
From Table \ref{KPM-Times}, we observe that the overall RTT incurred in \systemName for different quantized ML models (for xApp inference) is well below the $< 1$s timing requirement of O-RAN's Near-RT RIC. Here, the RTT comprises time consumed in the entire closed-loop control between RAN and RIC, i.e., encryption time (in data encryption microservice), model evaluation time (within xApp), KPM data collection (from RAN to RIC database via E2 interface), and control decision (from RIC to RAN). 
Note that encryption time and model evaluation time vary for different time windows as shown in Table \ref{KPM-Times}.

\paragraph{Data Privacy Analysis}
Our proposed \systemName framework ensures secure execution of the xApp's ML-based inference model without decrypting the sensitive data to preserve the RAN/user privacy. \systemName is designed to guard against the threats described in Section~\ref{Threat} as follows.

\smallskip \noindent $\bullet$ To ensure RAN/user data privacy, the data is first encrypted using IPFE in such a way that no entity who may be able to intercept this data, can decrypt it or even learn anything about the statistics because it is encrypted using a secret key which is needed for decryption, and the secret keys are securely kept private only to the encryptor (i.e., UE/RAN).       
        
\smallskip \noindent $\bullet$ In \systemName, the \xAppName uses the functional decryption key $\text{sk}_\mathbf{y}$ to get the output of the first layer without decrypting or learning the user/RAN data. Specifically, upon accessing the encrypted data, \systemName ensures that the \xAppName gives its inference without knowing the raw data. 

\smallskip \noindent $\bullet$ Given that the number of functional decryption keys distributed (which is a function of the number of neurons in the first hidden layer) is fewer than the number of inputs, it would be challenging to obtain the input private data as the number of linear equations that can be constructed using the model weights and the results of the inner product operations is less than the number of unknowns (input data). Also, the solution space is sufficiently large under this condition even after applying quantization. Specifically, given $n$ and $m$ are the number of inputs to the neural network and functional decryption keys distributed, respectively, linear algebra posits that the solution space of the plaintext data is $256^{n - m}$, which grows exponentially as the number of functional decryption keys decreases. 
This complexity makes it hard for attackers to infer the input, and to address this concern, the KDC controls the number of functional decryption keys generated, thereby providing a countermeasure against these attacks. 

\smallskip \noindent $\bullet$ To protect against unauthorized extraction of the input data, the KDC must ensure that the column space of matrix $\mathbf{W}$ does not include any standard basis vectors, i.e., vectors with a single component equal to 1 and all other components equal to 0. For example, consider the matrix $\mathbf{W}$ defined as:
$$\mathbf{W} = \begin{pmatrix}
1 & 0 \\
0 & 1 \\
0 & 0 \\
\end{pmatrix}$$ 
This matrix represents a neural network with 3 nodes in the input layer and 2 nodes in the first hidden layer. If the KDC distributes decryption keys corresponding to this matrix, a malicious xApp could determine two elements of the inputs. Thus, none of the standard basis vectors must lie in the column space of $W$ to prevent any part of the input from being deduced through linear equations. The necessary check can be efficiently implemented using Gaussian Elimination to ensure that $W$ is secure for use, preventing the malicious xApp from extracting any individual input components. 
   
\smallskip \noindent $\bullet$ Although \xAppName has access to $g^r$ and $h_i = g^s$, acquiring the secret key $\mathbf{s}$, the random number $r$, or even $g^{sr}$ and utilizing them to decrypt data $\mathbf{x}$ is highly challenging because it requires solving a hard discrete logarithmic problem as these values are elements in $Z_p$ with very large numbers. Moreover, having the functional decryption key $\text{sk}_\mathbf{y}$ and ciphertext ($g^r$ and $g^{(sr+x)}$), it is also computationally intractable to obtain the data $\mathbf{x}$ due to the impracticality of solving the discrete logarithm problem as the data is masked with $g^{s r}$ in the encryption process, as seen in Eq.~\ref{Enc_eq}.


\begin{table}[t]
    \centering
    \caption{\systemName Round Trip Time} 
    \vspace{-0.1in}
       \scriptsize
    \begin{tabular}{|p{0.85in}|p{0.5in}|p{0.6in}|p{0.6in}|p{0.5in}|}
        \hline
         Time Windows (Input Shape) & Encryption Time & Model Evaluation Time & Round Trip Time\\ \hline
         5 (25,1) & 0.01s & 0.046s & 0.109s \\ \hline
         10 (50,1) & 0.024s & 0.167s & 0.244s  \\ \hline
         20 (100,1) & 0.045s & 0.429s & 0.527s  \\ \hline
    \end{tabular}
    \label{KPM-Times}
    \vspace{-0.25in}
\end{table}

\textbf{Observation.} Based on our experiments, it is evident the \systemName framework unequivocally ensures user/RAN data privacy against diverse threat models, preserves xApp inference accuracy, and meets stringent O-RAN timing requirements. Consequently, we strongly endorse the adoption of \systemName in real-world O-RAN deployments, especially in DoD scenarios where zero-trust principles are indispensable.

\vspace{-0.12in}
\section{Conclusion and Future Work}
\vspace{-0.05in}
This paper introduces a novel privacy-preserving \systemName framework, which is based on a robust cryptographic  technique, called, utilizing inner product functional encryption (IPFE). Our IPFE-based \systemName enables efficient computation on encrypted data for ML-based xApps, ensuring the protection of sensitive RAN and UE information without compromising accuracy or O-RAN latency requirements. We conducted extensive experiments on an over-the-air O-RAN testbed, using \xAppName, to validate \systemName's effectiveness in preserving data privacy while still enhancing xApp performance and meeting latency thresholds. Additionally, a potential future work involves designing a privacy-preserving solution framework that eliminates the need for a trusted central entity like the Key Distribution Center (KDC).

\vspace{-0.1in}
\section{Acknowledgement}
\vspace{-0.05in}
The project is partially supported by the Public Wireless Supply Chain Innovation Fund under Federal Award ID Number 51-60-IF007.

\bibliographystyle{IEEEtran}
\bibliography{main}
\vspace{-0.2in}

\end{document}